\documentclass[aps,prl,floats,superscriptaddress,showpacs,twocolumn]{revtex4-1}

\usepackage{amsmath, amsthm, amssymb}
\usepackage{enumitem}
\usepackage{graphics,graphicx}
\usepackage{epsfig}
\usepackage{times}
\usepackage{dcolumn}
\usepackage{bm}

\usepackage{ulem, color}

\begin{document}

\title{Synchronization of interconnected networks: the role of connector nodes}

\author{J. Aguirre}
\affiliation{Centro de Astrobiolog\'{\i}a, CSIC-INTA. Ctra.\
de Ajalvir km 4, 28850 Torrej\'on de Ardoz, Madrid, Spain}
\affiliation{Grupo Interdisciplinar de Sistemas Complejos (GISC)}
\author{R. Sevilla-Escoboza}
\affiliation{Centro Universitario de los Lagos, Universidad de Guadalajara, Enrique D\'{i}az de Leon, Paseos de la Monta\~na, Lagos de Moreno, Jalisco 47460, Mexico}
\affiliation{Laboratory of Biological Networks, Center for Biomedical Technology, UPM, Pozuelo de Alarc\'{o}n, 28223 Madrid, Spain}
\author{R. Guti\'{e}rrez}
\affiliation{Department of Chemical Physics, The Weizmann Institute of Science, Rehovot 76100, Israel}
\author{D. Papo}
\affiliation{Group of Computational Systems Biology, Center for Biomedical Technology, UPM, Pozuelo de Alarc\'{o}n, 28223 Madrid, Spain}
\author{J. M. Buld\'u}
\affiliation{Laboratory of Biological Networks, Center for Biomedical Technology, UPM, Pozuelo de Alarc\'{o}n, 28223 Madrid, Spain}
\affiliation{Complex Systems Group, Universidad Rey Juan Carlos, 28933 M\'ostoles, Madrid, Spain}

\date{\today}

\begin{abstract}
In this Letter we identify the general rules that determine the synchronization properties of interconnected networks. 
We study analytically, numerically and experimentally how the degree of the nodes through which two networks are connected influences the ability of the whole system to synchronize. We show that connecting the high-degree (low-degree) nodes of each network turns out to be the most (least) effective strategy to achieve synchronization.
We find the functional relation between synchronizability and size for a given network-of-networks, and report the existence of the optimal connector link weights 
for the different interconnection strategies.
Finally, we perform an electronic experiment with two coupled star networks and conclude that the analytical results are indeed valid in the presence of noise and parameter mismatches.
\end{abstract}

\pacs{89.75.Hc,89.75.Fb}


\maketitle
Real networks often interact with other networks of similar or
different nature, forming what is known as Networks-of-Networks ($NoN$)  \cite{gao2012a}. By
considering a $NoN$, new perspectives in the understanding of classical
network phenomena, such as robustness \cite{buldyrev2010,gao2011,gao2012}, spreading \cite{dickinson2012,saumellmendiola2012} or interaction between modules \cite{radichi2013,martin2013}, can be obtained, sometimes
with counter-intuitive results. Similarly, while
synchronization in complex networks has been widely studied \cite{arenas2008}, very
few works have investigated synchronization in $NoN$s. Huang et al. \cite{huang2006}
showed that when two networks interact through random connections an
exact balance between the weight of internal links in a network and
the weight of links between networks results in greater
synchronization between the two networks. It has also been shown that
for multiple interacting networks, random connections between distant
networks increase the synchronization of the complete $NoN$ \cite{park2006}.

Real networks exhibit high heterogeneity of the node degree, with hubs (i.e., high-degree) and peripheral (i.e., low-degree) nodes \cite{newman2010}. What happens if connector links between the networks, termed
{\it inter-links}, are not randomly created, but are instead chosen
according to a particular connection strategy?
Carlson et al. \cite{Carlson2011} analyzed the influence that low-degree nodes may have  on the collective dynamics of networks.
 Wang et al. \cite{wang2006} showed that when two neuron clusters get connected, both the heterogeneity of the network
and the degree (i.e. number of connections) of the {\it connector nodes}, i.e. the nodes reached by inter-links, influence the coherent behavior of the whole system.
A recent study demonstrated that the proper selection of connector 
nodes has strong implications on structural (centrality) and dynamical properties (spreading or population dynamics) occurring in a $NoN$ \cite{aguirre2013}.

In this Letter,  we  study in a systematic way how connector nodes 
between a group of networks with heterogeneous topology affect synchronization and stability of the resulting $NoN$, and provide general rules for electing 
in a non-random fashion the connector nodes that maximize the synchronizability.

The stability of the synchronized state of a group of coupled identical dynamical units is given by the corresponding Master Stability Function (MSF) \cite{pecora1998}. 
For a given dynamical system and coupling form, the stability of synchronization depends on the second lowest eigenvalue $\lambda_2$,
usually called the {\it spectral gap} or {\it algebraic connectivity}, and the largest eigenvalue
$\lambda_N$ \cite{boccaletti2006} of the network Laplacian matrix ${\bf L}$ \cite{laplacian}.
 Dynamical systems can then be classified according to their MSF \cite{others}: a) class I systems never 
 synchronize irrespective of their network topology,
 b) class II systems synchronize for values of $\lambda_2$ above a threshold given by the MSF, and 
 c) class III systems synchronize 
for eigenratios $r=\lambda_N/\lambda_2$ lower than a  threshold determined by the MSF.

For isolated networks, the eigenratio $r$ has been used as an indicator of synchronizability both in theoretical 
\cite{barahona2002,zhao2006} and in real systems such as functional 
brain networks \cite{basset2006,schindler2008}. For class III systems, obtaining a maximally 
synchronizable system is tantamount to minimizing the eigenratio $r$.
Nishikawa et al. \cite{nishikawa2006} showed that when the network structure and the link weights were adequately transformed into unidirectional hierarchical organizations, 
the minimum eigenratio  $r=1$ (since $\lambda_2 \leq \lambda_3 \leq ... \leq\lambda_N$) was achieved.
 
Given a fixed number of nodes $N$ and links $L$ it is also possible to reduce $r$ using genetic algorithms obtaining the so called {\it entangled networks} 
\cite{donetti2005,donetti2006}, which are characterized by high homogeneity of the node degree,
shortest path and betweenness. 
Crucially, these results indicate that a good strategy to enhance the synchronizability of a network is to disconnect the network hubs and connect nodes with low degree.

\begin{figure}[t]
\vskip-0.8cm
\centering
\includegraphics[width=0.45\textwidth]{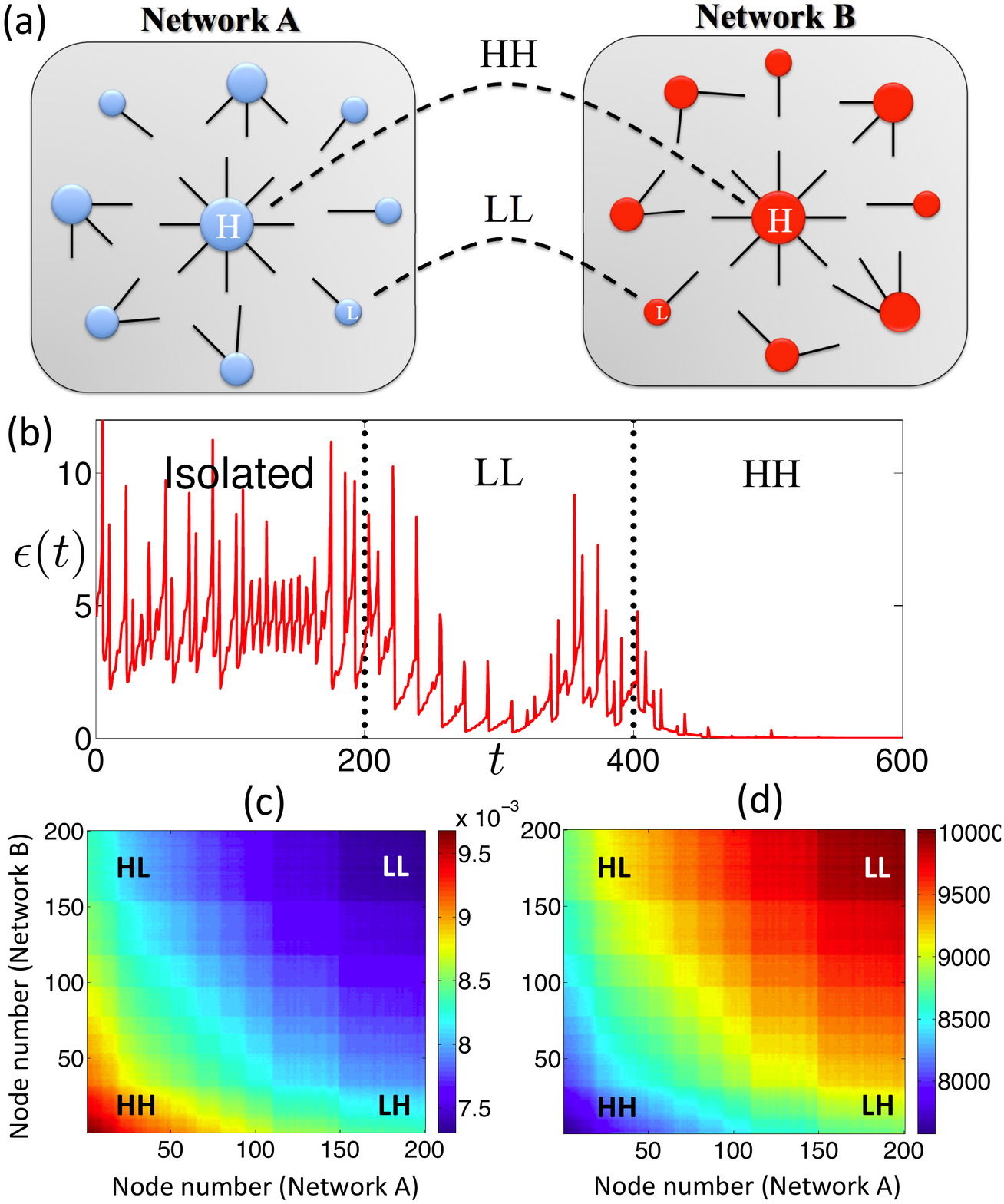}
\vskip-1.2cm
\caption{(Color online). (a) Schematic representation of the interconnection of two heterogeneous networks.
HH corresponds to a strategy connecting high-degree nodes and LL to the connection between low-degree nodes. (b) Synchronization error 
$\epsilon(t)$ of two interconnected Barab\'asi-Albert networks of $N=200$ R\"ossler oscillators at three different stages: 
isolated, interconnected following a LL strategy, and replacing 
the LL connection with a HH one. 
Equations of the R\"ossler system are given in \cite{rossler1976} and the parameters used in the simulations are $a=b=0.2$ and $c=5.7$.
$\epsilon (t)$ is obtained as the average across all pairs of oscillators of the pairwise distance in three dimensional phase space,
 $\frac{2}{N(N-1)} \sum_{i<j} \| {\bf x}_i - {\bf x}_j \|$, where  ${\bf x}_i$ is the state vector of oscillator $i$ and  $\| \cdot \|$ denotes the (Euclidean) norm.
(c) $\lambda_2$ of the $NoN$ obtained from connecting two $N=200$ Barab\'asi-Albert networks
 with one inter-link, in all possible configurations. 
The node numbers are ordered according to the node degree and, when coinciding, 
the eigenvector centrality.
(d) Eigenratio $r=\lambda_N/\lambda_2$ for the same case as (c). 
}
\label{fig01}
\vskip-0.3cm
\end{figure}

How to maximally synchronize two (or more) interconnected networks, on the other hand, is still poorly understood.
Figure \ref{fig01}(a) shows a qualitative example of two different types of connections between two networks,
{\it High-degree/High-degree} (HH) and {\it Low-degree/Low-degree} (LL). Figure \ref{fig01}(b) depicts the synchronization error $\epsilon (t)$ of two scale-free networks of R\"ossler oscillators \cite{rossler1976} coupled with different strategies.
The high $\epsilon (t)$ obtained when both networks are isolated decreases when a LL connection is created ($t=200$), but only
 goes to zero with a HH connection ($t=400$), indicating the attainment of complete synchronization.

The role of the connector nodes in synchronization can be quantified by their influence on the value of the eigenvalues $\lambda_2$ and the eigenratio $r$.
Figures \ref{fig01}(c-d) show the $\lambda_2$ and the $r$ of two Barab\'asi-Albert networks 
\cite{boccaletti2006} of $N=200$ nodes, 
 inter-connected with a unique link in all the $N^2$ possible configurations. As shown in Fig. \ref{fig01}(c), the region with the highest $\lambda_2$ turns out to be centered around the HH connections, 
while LL results in a lower $\lambda_2$, 
and the optimal strategy to connect networks of class II dynamical systems would be through their higher degree nodes. 
Regarding class III systems, Fig. \ref{fig01}(d) shows that the HH connection is the best option to reduce 
the eigenratio $r$ and increase the synchronizability of the $NoN$. 
Since isolated networks decrease their synchronizability when connecting their high-degree nodes \cite{donetti2005}, 
the results for inter-connected networks obtained in this Letter 
represent another important example of how the behavior of a single network may fundamentally differ from that of a $NoN$.

Recent studies \cite{martin2013,radichi2013} on class II 
synchronization of interdependent networks proved the existence of a phase transition in $\lambda_2$ after the addition of sufficient links,
obtaining powerful analytical results for general networks. However, the approximations made by the authors, as well as the strategies used
to connect the networks,
resulted in expressions that are not dependent on the degree of the connector nodes. For these reasons, those papers 
give no information on the influence the degree heterogeneity may have on the 
synchronizability of a network when the inter-links are selected according to different strategies. 
 
To obtain an analytical expression determining the influence of the connector nodes on the 
complete synchronization of a $NoN$, we consider one of the simplest $NoN$ showing some degree heterogeneity:  a system consisting of two star networks connected by one inter-link. 
Each star consists of $N$ nodes, one high-degree node (H) connected to $N-1$ low-degree nodes (L).
We call $w_{ij}$ the weight of the link connecting nodes $i$ and $j$ and, without loss of generality, we consider all links inside each star 
(i.e. {\it intra-links}) to have the same weight $w_{intra}$. We then connect the two stars through a single 
inter-link of weight $w_{inter}=a w_{intra}$ according to
three different strategies: HH, LL and HL (note that, due to the symmetry of the system, LH is equivalent to HL). 

\begin{figure}[b]
\centering
\vskip0.5cm
\includegraphics[width=0.47\textwidth]{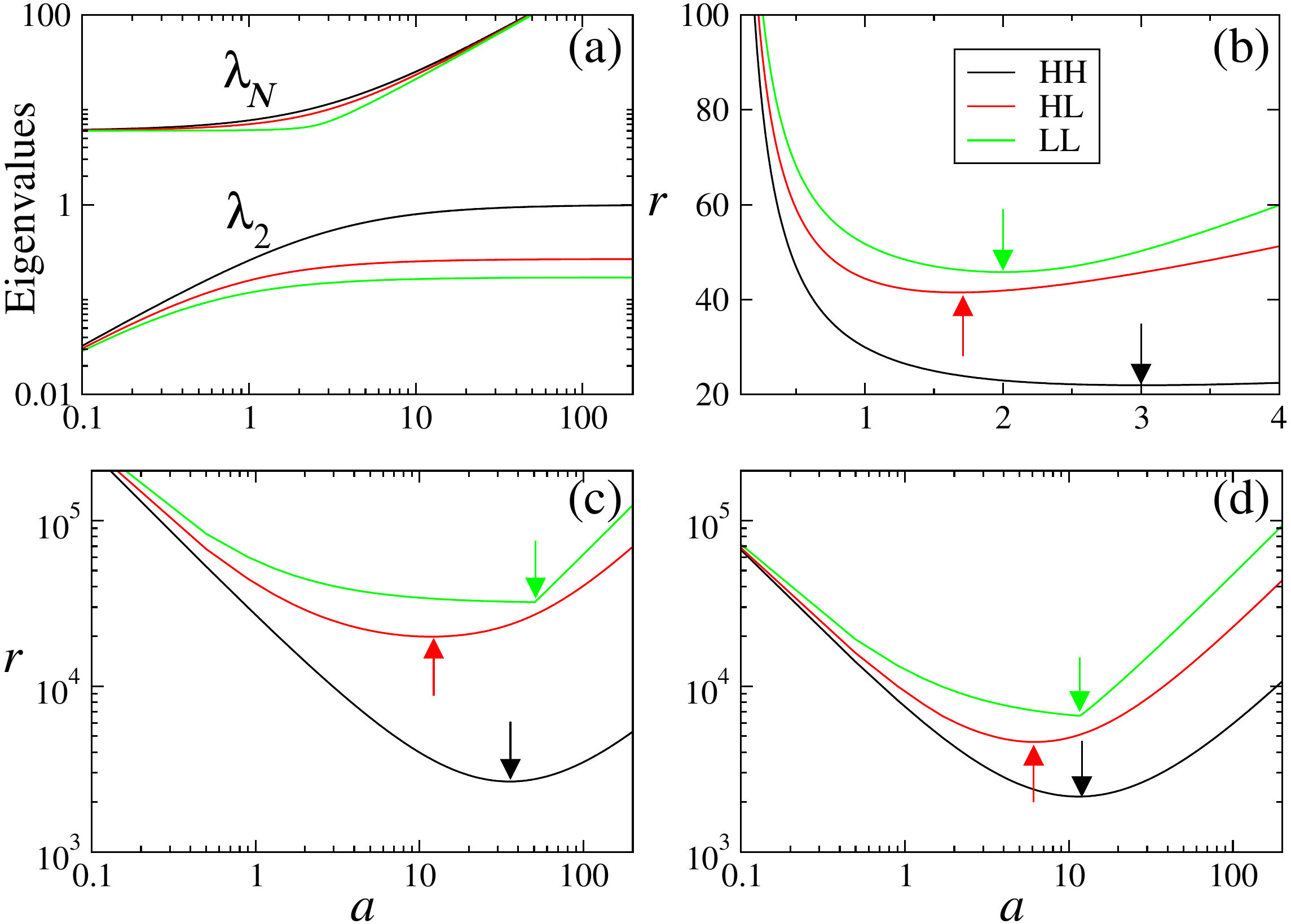}
\vskip-0.4cm
\caption{(Color online). Synchronizability for two networks connected by a single inter-link of weight $a$. 
(a) $\lambda_2$ and $\lambda_N$   
for two star networks of 6 nodes each. (b-c-d) Eigenratio $r$ for: (b) two star networks ($N=6$), 
(c) two scale-free networks ($N=500$), and (d) two Erd\H{o}s-R\'enyi random networks ($N=500$).  
Three connecting strategies are shown: HH (black), HL (red), and LL (green). 
The minima of the curves (arrows) correspond to maximum synchronizability \cite{kinks}. Plots (a-b) were obtained analytically
and (c-d) numerically.}
\label{fig02}
\end{figure}

Complete synchronization depends on the eigenvalues of the (weighted) Laplacian matrix. 
The symmetry of the configuration allows to reduce the characteristic polynomial of the Laplacian matrix so that $\lambda_N$ and $\lambda_2$ associated with the HH, 
LL and HL strategies
are respectively, and for all $N$ and $a$, the maximum and minimum roots of 
\vskip-0.4cm
\begin{equation}
x^3+C_2x^2+C_1x+C_0=0\,,
\label{Eq:lambdaN2total}
\end{equation}

\noindent where $C_2=-(1+N+2a)$, $C_1=N+4a+\xi a(N-2)$ and $C_0=-2a$, while $\xi$ is, depending on the connection strategy, $\xi^{\mathrm{HH}}=0$, $\xi^{\mathrm{HL}}=1$  or $\xi^{\mathrm{LL}}=2$ \cite{superindices} 
(see Section S1 of \cite{SM} for more details).
Without any loss of generality we have assumed that $w_{intra}=1$, thus being $w_{inter}=a$. The eigenvalues follow
\begin{equation}
\lambda_{N,2}=N/2+a \pm \sqrt{(N/2)^2+a^2+(N-2)a}
\label{Eq:lambdaN2CC}
\end{equation}
for the HH connection, while for the LL and HL strategies they take a more complex analytical form (see Fig. \ref{fig02}(a) and Table S1 of \cite{SM}). 

For networks composed of class II systems, Eq. \ref{Eq:lambdaN2CC} yields that increasing the weight of the inter-link $a$ 
increases $\lambda_2$ (decreasing the network modularity \cite{fortunato2010}) and therefore the synchronizability of the $NoN$.
The same behavior characterizes the LL and
HL strategies (see Fig. \ref{fig02}(a) and Table S2 of \cite{SM} for details). Furthermore, comparing Eq. \ref{Eq:lambdaN2CC} with those obtained in the LL and HL strategies shows that
$\lambda_2^{\mathrm{HH}}>\lambda_2^{\mathrm{HL}}>\lambda_2^{\mathrm{LL}}$  for the meaningful values of $N$ and $a$, that is, $N>2$ and $a>0$. 
Thus, for class II systems the optimal strategy is always the one connecting high-degree nodes.

Next, we can investigate which of the strategies leads to the lowest eigenratio $r$ in class III systems. 
The totally algebraic solution of the two-star system  allows us to prove that, 
for all feasible values of $N$ and $a$, $r^{\mathrm{HH}}<r^{\mathrm{HL}}<r^{\mathrm{LL}}$ (see Section S1.3 of \cite{SM} for the details). Thus, HH turns out to be the strategy optimizing synchronizability of the $NoN$. 
Figures \ref{fig02}(b-d) show the evolution of $r$ as a function of the inter-link weight $a$ 
for the three connecting strategies and for different network topologies. 
Even though no closed analytical expression can be found for complex topologies, the Laplacian of such networks can be studied numerically, leading to the same conclusions in complex networks.
In all cases, the HH type of
connection leads to the lowest $r$, suggesting that the results proved for the two star system 
are of general applicability. 

Importantly, class III systems have an optimal inter-link weight $a_{sync}$ minimizing $r$.
This fact is easy to verify in the case of two star networks, because for all connecting strategies, 
$\lim_{a \to \infty} r=\lim_{a \to 0} r=\infty~\forall~ N>2$. Furthermore, it is worth noting 
that the optimal inter-link weights $a_{sync}^{\mathrm{HH}}$, $a_{sync}^{\mathrm{HL}}$ and $a_{sync}^{\mathrm{LL}}$ do not coincide (see the arrows in Fig. \ref{fig02}(b-d) and Section S1.3 of \cite{SM} for the analytical details).

\begin{figure}[t]
\centering
\includegraphics[width=0.55\textwidth]{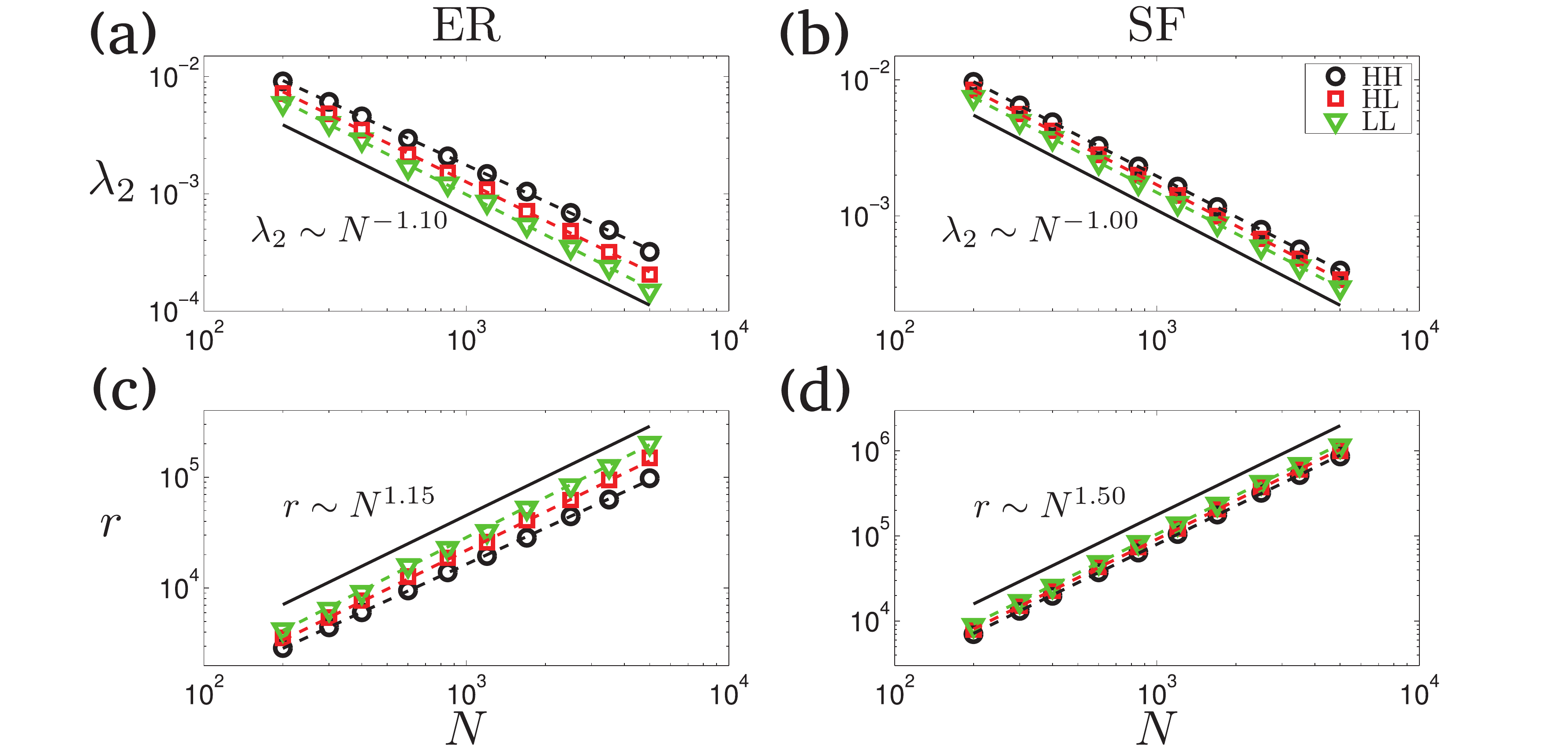}
\vskip-0.2cm
\caption{(Color online). Dependence of synchronizability of class II and III systems on the size of the networks $N$. 
Averaged second eigenvalue $\lambda_2$ (a and b) and $r$ (c and d) over 30 realizations of two Erd\H{o}s-R\'enyi networks and two scale-free networks of average degree $\bar k=12$.  See Section S2 of \cite{SM} for details.
}
\label{figSvsN}
\vskip-0.6cm
\end{figure}

To conclude the analytic study of the problem, we note that increasing the number of nodes always hinders synchronizability,
as indicated by $d\lambda_2/dN<0 ~\forall~a>0$ for class II systems and $dr/dN>0 ~\forall~a>0$ for class III systems. 
This result goes beyond two star networks and is
valid for networks of more complex topology. In Fig. \ref{figSvsN} we can observe how the scaling
of synchronizability with $N$ changes according to the topology of the networks (see S2 of \cite{SM} for some theoretical arguments lending support to these results).

We now prove the robustness of our results with a network of electronic circuits. 
The experimental setup consists of two diffusively coupled
 star networks of piecewise R\"{o}ssler circuits \cite{rossler1976,CP} operating in a chaotic regime (see Section S4 of \cite{SM} for details of the electronic circuits) 
 \cite{Pisarchik06}. They follow the same two-star topology described above, with both $w_{intra}$ and $w_{inter}$ as experimentally accessible parameters. 
 

\begin{figure}[t]
\vskip-0.3cm
\centering
\includegraphics[width=0.52\textwidth]{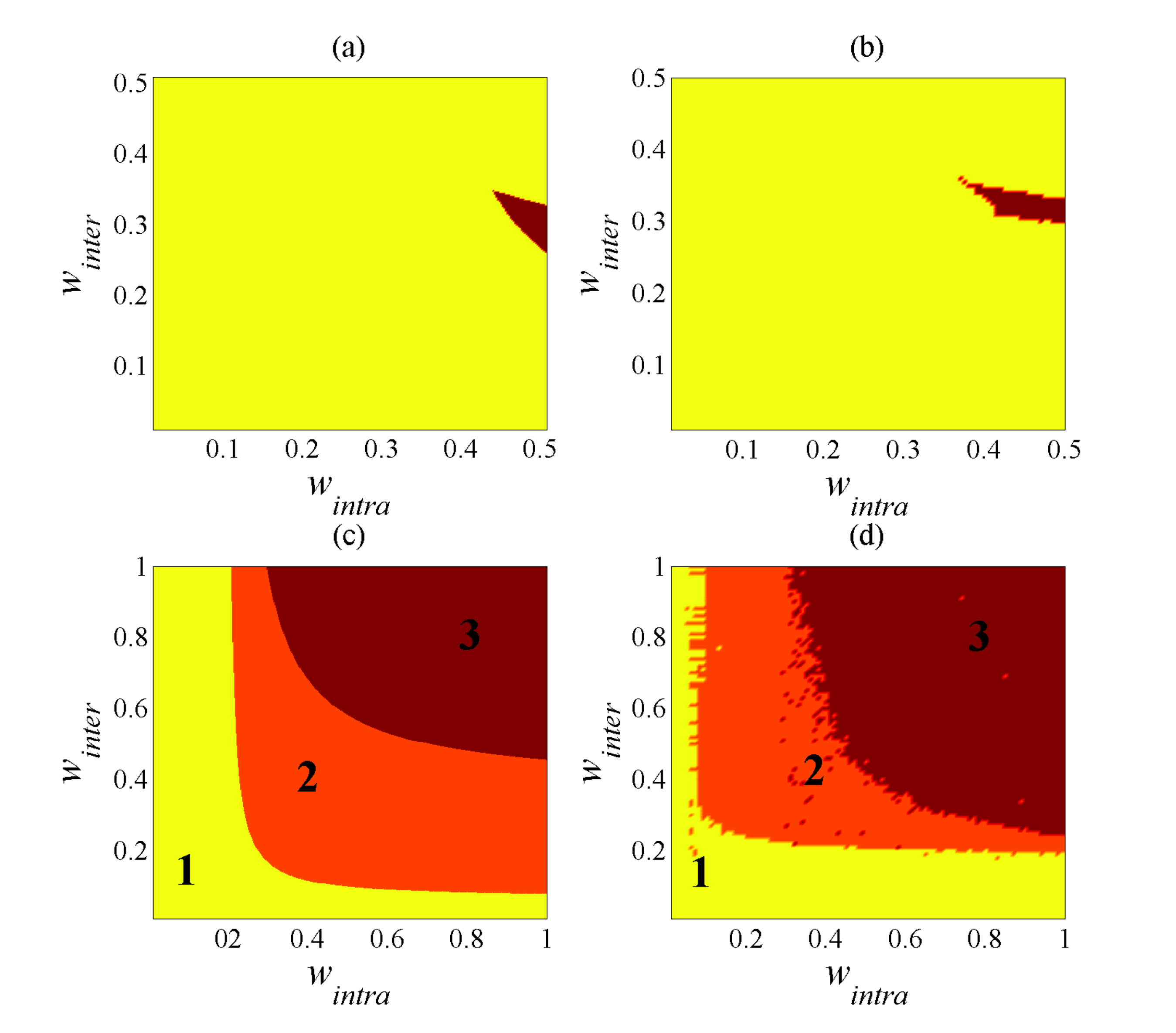}
\vskip-0.5cm
\caption{(Color online). Experimental verification of the phenomenology presented here. (a) and (b) show the regions of complete synchronization of two star networks of type III 
R\"ossler systems coupled by a HH strategy. Neither HL (LH) 
nor LL strategies lead to synchronization, as predicted 
by the theory and confirmed by the experiments (not shown here). (c) and (d) depict class II R\"ossler systems. 
Regions correspond to: 1) no synchronization, 
2) complete synchronization with the HH strategy and 3) complete synchronization with the HH and the LL strategies. 
Results are theoretical (a and c) and experimental (b and d). 
The zeroes of the MSF are $\nu_1=0.107$ and $\nu_2=2.863$ for class III and $\nu_c=0.0651$ for class II. 
}
\label{fig04}
\vskip-0.5cm
\end{figure}

It is important at this stage to recall that while maximizing $\lambda_2$ (in class II) and minimizing $r$ 
(in class III) increase the synchronizability of a network, it is 
the MSF that ultimately determines if complete synchronization is achieved \cite{pecora1998}. Coupling through the $x$ variable leads to a class III system of equations.
For class III systems, the zeroes of the MSF ($\nu_1$ and $\nu_2$)
 determine the synchronization region, where the network has to fulfill the conditions $\sigma\lambda_2>\nu_1$ 
 and $\sigma \lambda_N<\nu_2$, where $\sigma$ is the coupling strength. 
The theoretical treatment of the class III R\"ossler systems described in Sections S4 and S5 of \cite{SM} indicates that
the HH strategy is the only one fulfilling the former requirements given by the MSF. For this case, Fig. \ref{fig04} shows 
qualitatively similar results for the synchronization regions in the ($w_{intra},w_{inter}$) phase space obtained 
theoretically (a) and experimentally (b) \cite{diff_exp}. 
In the latter, the synchronization region is determined by computing the average of the synchronization error  $\langle\epsilon\rangle$ 
of all units of the $NoN$, where the error between systems $i$ and $j$ is given by $\epsilon_{i,j}=\lim_{T\to \infty} T^{-1} \int_0^T \lVert x_i(t) - x_j(t) \rVert dt$ \cite{sync_err}. 

When the coupling is introduced through  $y$, the systems become of class II  \cite{huang2009}. 
In this case, the MSF only has one zero $\nu_c$ 
and synchronization only requires $\sigma\lambda_2>\nu_c$. Figure \ref{fig04}(c) depicts the synchronization regions obtained theoretically for different connecting strategies. 
The HH strategy turns out to require less internal and external coupling. 
Qualitatively similar results were obtained experimentally, as shown in Fig. \ref{fig04}(d).



In conclusion, in this work we  showed that whenever 
two networks are connected by one inter-link, 
the degree of the connector nodes plays a fundamental role in achieving synchronization.
Connecting high-degree nodes is, by default, the best synchronization strategy, 
while connecting low-degree nodes is the worst option. 
Interestingly, increasing the number of inter-links leads to the same qualitative results (see Section S3 of \cite{SM} for details). Furthermore, synchronizability always decreases as a power-law of the size of the system for both classes.
On the other hand, while increasing 
the inter-link weight consistently favors complete synchronization for class II systems, 
for class III there is an optimum 
value of the inter-link weight that depends on the connecting strategy. 
Our results are generic and independent of the size or topology of the networks, as indicated by numerical simulations of networks with more complex topologies (e.g., ER random or scale-free).


Possible applications of our methodology could be the design of optimal interconnection strategies in groups of interacting networks,
such as power grids \cite{motter2013} and {\it ad hoc} mobile networks \cite{diazguilera2008}, or the identification of the links to be deleted
in processes where high synchronizability plays against the normal functioning of the system (such as in epilepsy \cite{schindler2008}).

Authors acknowledge J.A. Capit\'an, D. Hochberg and M.A. Mu\~noz for fruitful conversations and their careful reading of the manuscript, and the support of MINECO 
(FIS2011-27569 and FIS2012-38949-C03-01) and of CAM (MODELICO-CM S2009ESP-1691). R.S.E. acknowledges UdG, Culagos (Mexico) for financial support (PIFI 522943 (2012)
 and Becas Movilidad 290674-CVU-386032).


\begin{thebibliography}{}
\bibitem{kinks} The sharp changes in the slope of the LL curves in (c) and (d) are due to a contingent approach between
eigenvalues $\lambda_N$ and $\lambda_{N-1}$ that has no direct relation with the phenomenology presented here.
\bibitem{sync_err} The threshold for considering complete synchronization of the whole network $\epsilon_{thres}=0.096$ is obtained from the synchronization error of 
two identical unidirectionally coupled R\"ossler systems in the limit of large coupling.
\end{thebibliography}
\end{document}